\documentclass[]{spie}  

\usepackage{amsmath,amsfonts,amssymb}
\usepackage{graphicx}
\usepackage{mathtools}
\usepackage[colorlinks=true, allcolors=blue]{hyperref}

\title{Proper evaluation of spatially correlated noise in interferometric images}

\author[a,b,c,d]{Takafumi Tsukui}
\author[c,d]{Satoru Iguchi}
\author[c,e]{Ikki Mitsuhashi}
\author[c,d]{Kenichi Tadaki}

\affil[a]{Research School of Astronomy and Astrophysics, Australian National University, Cotter Road, Weston Creek, ACT 2611, Australia}

\affil[b]{ARC Centre of Excellence for All Sky Astrophysics in 3 Dimensions (ASTRO 3D)}

\affil[c]{National Astronomical Observatory of Japan, National Institute of Natural Sciences, 2-21-1 Osawa, Mitaka, Tokyo, Japan.}

\affil[d]{Department of Astronomical Science, SOKENDAI (The Graduate University for Advanced
Studies)\\ 2-21-1 Osawa, Mitaka, Tokyo, Japan.}

\affil[e]{Department of Astronomy, The University of Tokyo, 7-3-1 Hongo, Bunkyo, Tokyo 113-0033, Japan}

\authorinfo{Send correspondence to T.T. E-mail: tsukuitk23@gmail.com}

\begin{document} 
\maketitle

\begin{abstract}

Recent interferometers (e.g. ALMA and NOEMA) allow us to obtain the detailed brightness distribution of the astronomical sources in 3 dimension (R.A., Dec., frequency). However, the interpixel correlation of the noise due to the limited uv coverage makes it difficult to evaluate the statistical uncertainty of the measured quantities and the statistical significance of the obtained results. The noise correlation properties are characterized by the noise autocorrelation function (ACF). We will present the method for (1) estimating the statistical uncertainty due to the correlated noise in the spatially integrated flux and spectra directly from the noise ACF and (2) simulating the correlated noise to perform a Monte Carlo simulation in image analyses. Our method has potential applications to a range of astronomical images of not only interferometers but also single dish mapping observation and interpolated and resampled optical images.
\end{abstract}


\section{INTRODUCTION}
\label{sec:intro}  
%

Recent developments in large interferometers (e.g. ALMA and NOEMA) have made it possible to spatially resolve the brightness distribution of astronomical objects. These observations have enabled us to obtain a three-dimensional (R.A., Dec., and the line-of-sight velocity) structure of the gas emission and two-dimensional images of the continuum with high spatial resolution and sensitivity. As a result, the data allow for detailed image analysis; for example, spectral features of spatially resolved regions, faint and extended structures, and Fourier analysis of the image, etc. However, the interpixel correlation of the noise in interferometric images makes it difficult to evaluate the uncertainty of the results. There has been a lack of quantitative understanding of the interpixel correlation of noise and methods to evaluate the statistical uncertainty of the measured quantities and the significance of the scientific result, such as signal detection and image analysis (see the analytical consideration of the effect of spatially correlated noise on parameter estimation \cite{Davis2017-lu, Refregier1998-ui}). In this paper, we present the method and the Python code to characterize the spatial correlation of noise by measuring the noise autocorrelation function (ACF) and evaluate its effect on the measured quantities and the analysis results by using the measured noise ACF.

This paper is organized as follows.
In Section \ref{sec:method}, we present the noise correlation properties of ALMA data characterized by the autocorrelation function (ACF) and show that the noise correlation originates from the synthesized beam (dirty beam) structures, which remain even in the clean image and cannot be removed by any deconvolution algorithm. In Section \ref{sec:result2}, we also introduce the method (1) for estimating the statistical uncertainties associated with spatially integrated flux or spectra and (2) for generating the simulated noise maps from the measured noise ACF, which are useful to estimate the statistical significance of the result obtained by any image analysis, with example applications to real scientific data from Tsukui and Iguchi 2021\cite{Tsukui2021-sg}.

Throughout the paper, we use the noise map from emission line cube and continuum image data taken by ALMA Band 7 (2017.1.00394.S PI=González López, Jorge) as an example, but the method proposed by this paper can be applied not only to ALMA data but also to a variety of astronomical images (spatial correlation of the noise is expected in, e.g., images by the on-the-fly mapping with a single dish radio telescope with complex beam response and optical images which are resampled and interpolated). The code is publicly available from \url{https://github.com/takafumi291/ESSENCE}. 

\section{NOISE CHARACTERIZATION OF INTERFEROMETRIC IMAGE}

\label{sec:method}  
\subsection{The characterization of spatially correlated Noise}
\label{subsec:acf}
\begin{figure}[h]
\centering
\includegraphics[width=0.5\textwidth]{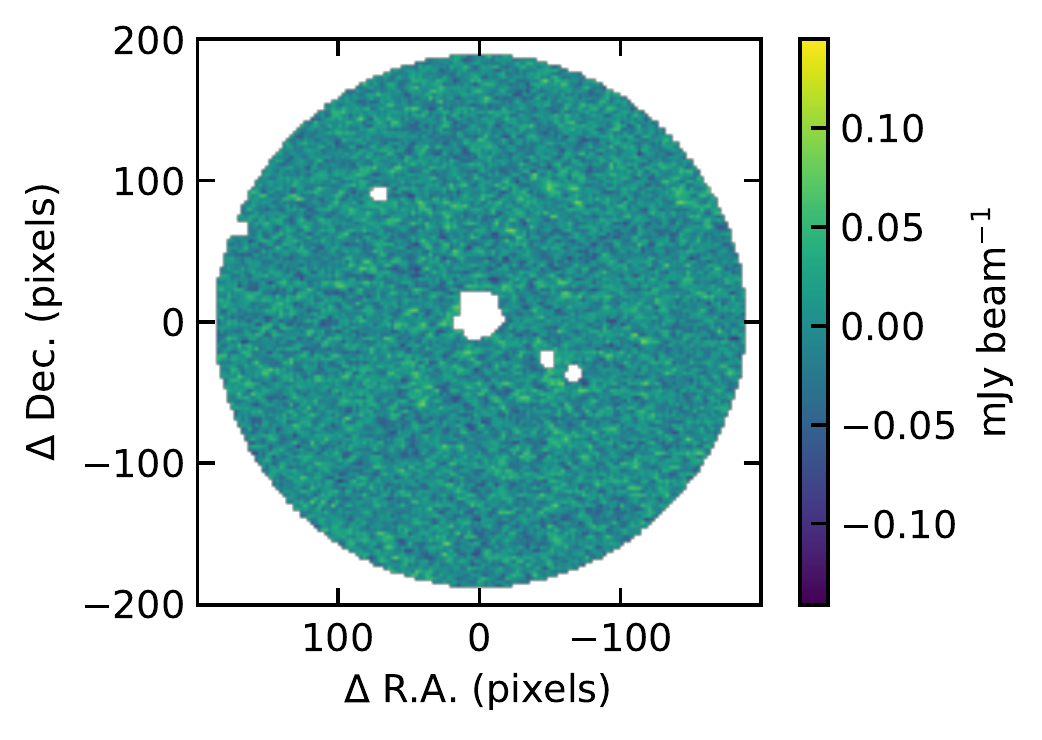}
\includegraphics[width=0.37\textwidth]{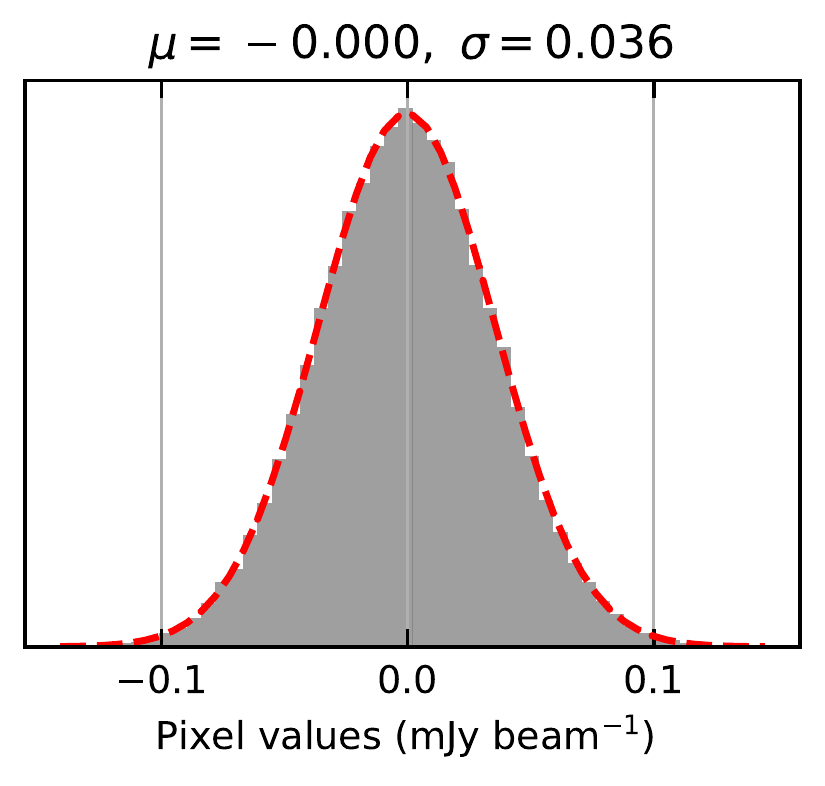}
\caption{Left: Example ALMA band 7 noise map. The source emission region is eliminated with the iterative 4 sigma clipping, see text. Right: The histogram of the pixel values of the noise map. \label{fig:fig1}} 
\end{figure}

First, we consider a two-dimensional noise map $N(\mathbf{x})$, where $\mathbf{x}$ denotes the position of the pixels, and pixel regions with signals from the object of interest are excluded. The statistical properties of the noise are assumed to be uniform in the noise image, which appears to be valid in the interferometric image\footnote{We discuss this in Section \ref{subsec:origin}}.  
In most of the literature, the noise in the radio interferometric image is quantified and reported with the root mean square of the noise map $N(\mathbf{x})$,
\begin{equation}\label{eq:eq1}
\begin{split}
     \sqrt{\langle N(\mathbf{x})^2\rangle} & \equiv \sigma_\mathrm{N},   \\
\end{split}
\end{equation}
where the brackets denote the expected value for each pixel, which is practically estimated by averaging over the noise maps. The mean of the noise in the image is nearly zero $\langle N(\mathbf{x})\rangle \equiv \mu \approx 0$, since most of the noise represented by the system temperature, $T_{\mathrm{sys}}$, is uncorrelated in a pair of antennas and, therefore, the power of the noise does not appear in the correlator output of interferometers such as ALMA. Also, extended background emission such as the cosmic microwave background (CMB) is resolved out without total power observation. However, these noises contribute to the random noise associated with visibility measurements, which propagate into the noise on the image by the Fourier transform. In the rest of the paper, we assume the mean of the noise map to be zero or already have been subtracted in other cases, and thus the root mean square and the standard deviation of the noise can be used interchangeably. Figure~\ref{fig:fig1} shows the example ALMA band 7 noise map and its histogram from the observation targeting the hyper luminous infrared galaxy BRI 1335-0417 at redshift of 4.4, which will be used in the later Section. The noise map is created by (1) eliminating the source of interest using 4 sigma clipping and the surrounding pixels around the clipped pixels with 3 full width of half maximum (FWHM) of the synthesized beam, and (2) using the image region where the primaly beam attenuation is less than 50\%.

When the noise can be assumed as Gaussian, the statistical and correlation properties of the noise are fully quantified by the noise autocorrelation function (ACF)\cite{Refregier1998-ui},
\begin{equation}\label{eq:eq2}
\xi(\mathbf{x}_{i,j})\equiv\langle N(\mathbf{x}+\mathbf{x}_{i,j})N(\mathbf{x})\rangle,
\end{equation}
where the expected value is estimated by averaging for all pixel pairs with the relative distance $\mathbf{x}_{i,j}=(i,j)$ in the noise image. The value of the ACF noise at zero lag, $\mathbf{x}_{i,j}=\mathbf{0}$, is equal to the variance of the noise as $\xi(0)= \langle N(\mathbf{x})^2\rangle =\sigma^2_N$. When the noise has no inter-pixel correlations, the noise ACF becomes
\begin{equation}\label{eq:eq3}
       \xi(\mathbf{x}_{i,j})= 
\begin{dcases}
    \sigma^2_N     & \text{if } \mathbf{x}_{i,j}=0\\
    0,             & \text{otherwise}
\end{dcases}
\end{equation}

In order to evaluate the statistical uncertainty of the derived noise ACF, we first considered the number of independent pixel pairs $N_{\mathrm{pair}}$ in the number of all available pairs $N'_{pair}$ used to evaluate the bracket in the Eq.~\ref{eq:eq2}, since the pixels within a beam are expected to be strongly correlated and not independent. We estimated the number of independent pixel pairs $N_{\mathrm{pair}}$ as the ratio of the number of all pixel pairs $N'_{\mathrm{pair}}$ and the number of pixels in the beam (beamarea in pixels)\footnote{The beam area in pixels is typically estimated by $2\pi b_\mathrm{maj} b_\mathrm{min}/8\mathrm{ln}2$, where $b_\mathrm{maj}$ and $b_\mathrm{min}$ are the major and minor FWHMs of the mainlobe of the synthesized beam ("clean" beam).} $A_{\mathrm{beam}}$,
\begin{equation}\label{eq:eq4}
     N_{\mathrm{pair}}=N'_{\mathrm{pair}}/A_{\mathrm{beam}}.
\end{equation}
Then, The associated statistical uncertainty of the noise ACF $\Delta\xi(\mathbf{x}_{i,j})$ is calculated as the usual standard error of the mean but with an independent sample size $N_{\mathrm{pair}}$, that is, the standard deviation of the multiplication of the values across all pairs of pixels with a separation $\mathbf{x}_{i,j}$ divided by the root of the number of independent pixel pairs $N_{\mathrm{pair}}$, 
\begin{equation}\label{eq:eq5}
     \Delta\xi(\mathbf{x}_{i,j})=\sqrt{\langle N(\mathbf{x}+\mathbf{x}_{i,j})^2 N(\mathbf{x})^2 \rangle/N_{\mathrm{pair}}}.
\end{equation}

\begin{figure}[h]
\centering
\includegraphics[width=0.43\textwidth]{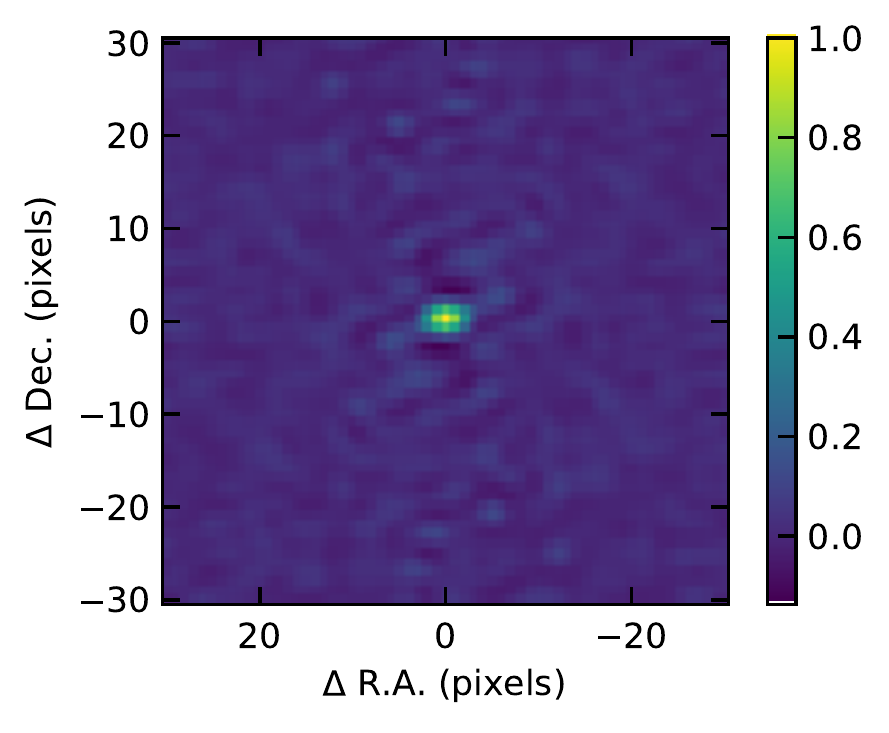}
\includegraphics[width=0.43\textwidth]{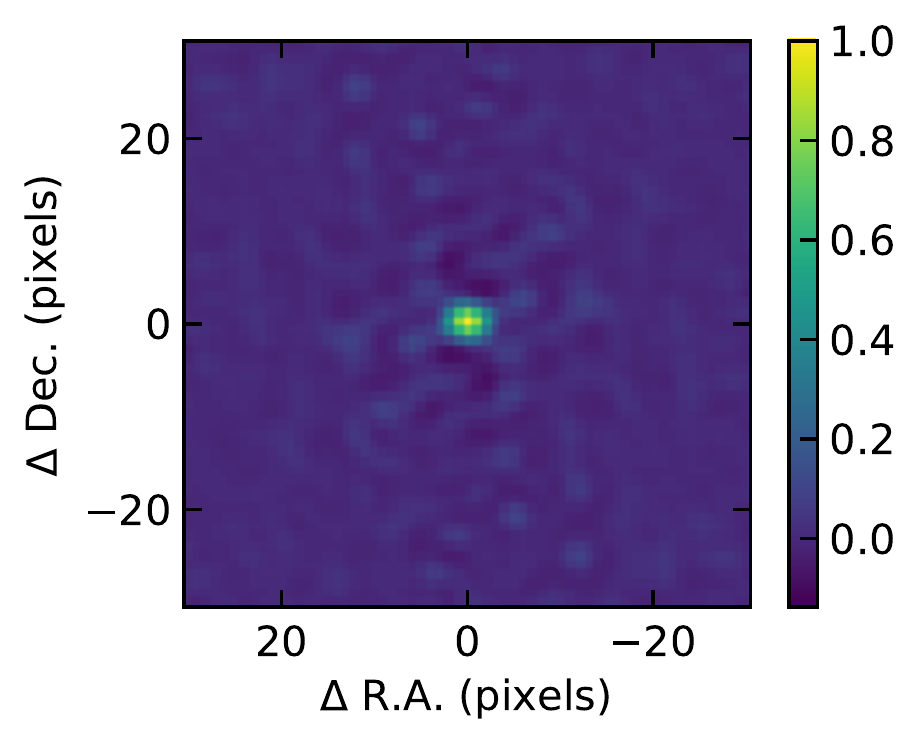}
\caption{The noise ACF computed for the ALMA band7 noise map (left), showing a similar pattern in the synthesized beam of the observation (right). \label{fig:fig2}} 
\end{figure}

Figure~\ref{fig:fig2} shows the results of the noise ACF (Eq.~\ref{eq:eq2}) computed for the noise map shown in Fig.~\ref{fig:fig1}, and the synthesized beam of the observation, both of which are normalized so that the central value is one. The noise ACF shows a pattern similar to that of the synthesized beam, with the strong short-range correlations near the center and long-range correlations away from the center corresponding to the main-lobe and side-lobe of the synthesized beam, respectively. This suggests that most of the correlation of the noise originates from the Fourier transform involved in the interferometric imaging with the spatial frequency filter, the limited spatial frequency coverage of the observation, which is illustrated in the following subsection.

Note that the noise ACF are measured for the noise maps of band 7 continuum image (shown in Figure~\ref{fig:fig2}), [C\textsc{ii}] line (velocity integrated over the velocity range of -400 to 400 km s$^{-1}$) moment 0 map, and the [C\textsc{ii}] line cubes (each velocity channel map) for later section. These maps are primary beam uncorrected and \textsc{clean}ed images produced by the \textsc{clean} algorithm in CASA (see details in \cite{Tsukui2021-sg}). 


\subsection{Origin of the noise correlation}
\label{subsec:origin}
In interferometric observations, the measurable quantities are visibility, Fourier amplitude, and phase coefficients of the astronomical image at the given spatial frequencies $(u,v)=\textbf{D}/\lambda$, which are related to the antenna baseline vector \textbf{D}\footnote{separation vector of pairs of antennas} projected onto the plane of the sky and the observed wavelength $\lambda$. The image is then computed by the Fourier transform of the measured visibility.

In order to explore the origin of the noise correlation in the image, seen in (Fig.~\ref{fig:fig2}), we start with the ideal case in which the observation measures visibility at all spatial frequencies ($u,v$). The visibility of the source of interest is $V(u, v)$, which is the Fourier transform of the true flux distribution of the source in the image, $\hat{S}(x,y)=\mathrm{FT}[V(u, v)]$, where FT denotes the Fourier transform. The measurement of $V(u,v)$ usually involves uncorrelated random noise, which we describe with the random variable $\hat{N}_{\mathrm{vis}}(u,v)$ with zero mean. We assume that the statistical property of the random variable $\hat{N}_{\mathrm{vis}}(u,v)$ is uniform as a function of $u$ and $v$, that is, the system noise temperature is the same for all antennas.
The image obtained $I(x,y)$ is the Fourier transform of the measurement $V(u,v)+\hat{N}_{\mathrm{vis}}(u,v)$,
\begin{equation}
\begin{split}
    I(x,y)&=\hat{S}(x,y)+\hat{N}(x, y)\\
          &=\mathrm{FT}[V(u,v)+\hat{N}_{\mathrm{vis}}(u,v)]  
\end{split}
\end{equation}
where $\hat{N}(x, y)=\mathrm{FT}(\hat{N}_{\mathrm{vis}}(u,v))$ is the noise component in the image, which is a random variable with zero mean\footnote{Fourier transform of the random variable with zero mean is also random variable with zero mean.}. 
The noise component of the image, $\hat{N}(x,y)$ is due to the random noise associated with the visibility measurement $\hat{N}_{\mathrm{vis}}(u,v)$, and the resulted noise map $N(x,y)=\hat{N}(x,y)$ is uncorrelated in the ideal case where all spatial frequencies are measured. 

In a practical observation, only visibility is measured at the limited spatial frequency \{($u_1,v_1$), ($u_2,v_2$),..., ($u_M,v_M$)\} (uv coverage). 
The spatial transfer function $W(u,v)$, defined as the visibility at spatial frequency $(u,v)$ is measured if $W(u,v)$ is non-zero, is described as 
\begin{equation}
    W(u, v)=\sum_{i=0}^M \delta(u-u_i,v-v_i)+\delta(u+u_i,v+v_i),
\end{equation} 
where $\delta$ is the Dirac delta function.
The synthesized beam $b(x,y)$ is the Fourier transform of the spatial transfer function $W(u, v)$, $b(x,y)=\mathrm{FT}[W(u,v)]$. The resulting image $I(x,y)$, decomposed as the signal $S(x,y)$ from the source and noise map $N(x,y)$, is 
\begin{equation}
\begin{split}
    I(x,y) & =S(x,y)+N(x,y)\\
           & =\hat{S}(x,y)*b(x,y)+\hat{N}(x,y)*b(x,y)\\
           & =\mathrm{FT}[(V(u,v)+\hat{N}_{\mathrm{vis}}(u,v))W(u,v)],
\end{split}
\label{eq:eq8}
\end{equation}
where $*$ represents convolution.
As the noise correlation pattern (noise ACF) and the synthesized beam show a similar pattern in Fig. 2, the noise component of the image $N(x,y)$ is the convolution product of the random variable $\hat{N}(x,y)$ and the synthesized beam $b(x,y)$. This fact makes the noise in the image behave well; in particular, its statistical properties are uniform over the image, as we assumed to measure the noise ACF in Section \ref{subsec:acf}. 

\begin{figure}[h]
\centering
\includegraphics[width=0.43\textwidth]{NOISE_ACF.pdf}
\includegraphics[width=0.43\textwidth]{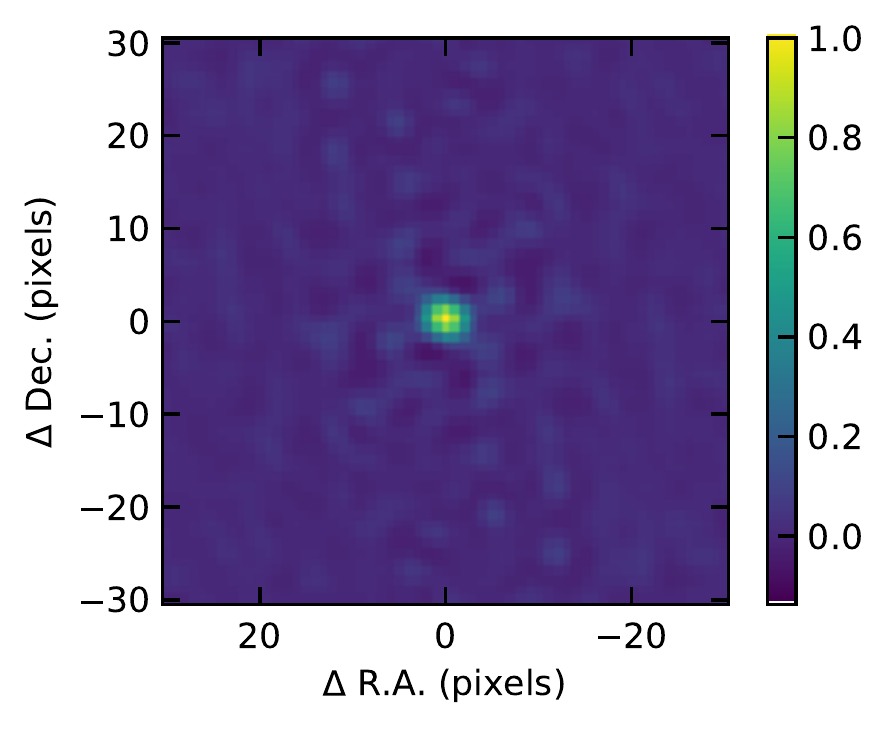}
\includegraphics[width=0.43\textwidth]{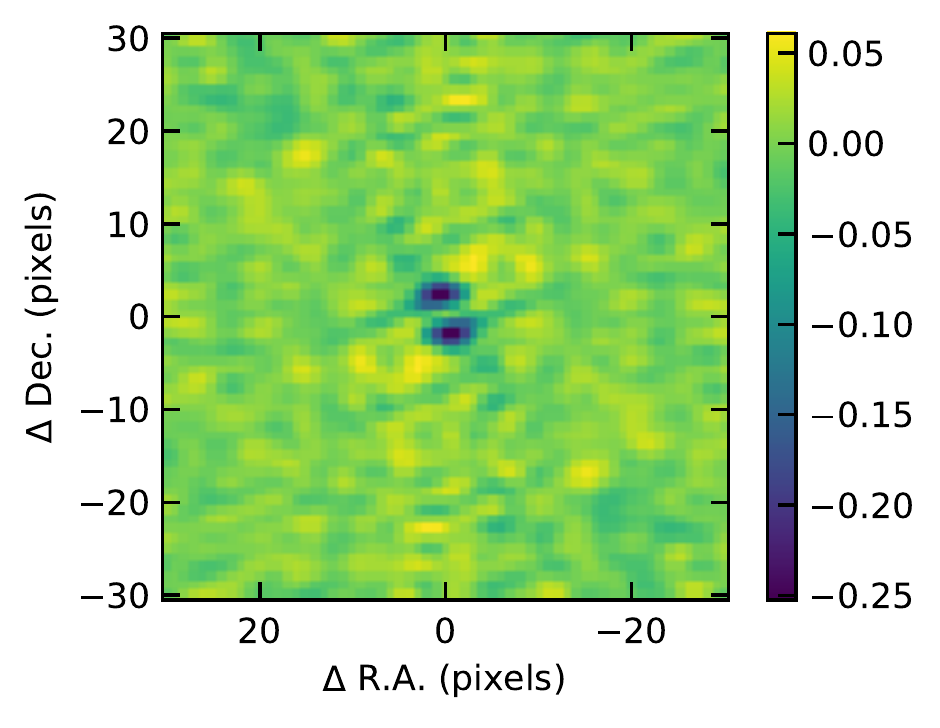}
\caption{Top left: the same noise ACF shown in Fig.~\ref{fig:fig2}, Top right: the ACF of the synthesized beam, Bottom: The residual of the noise ACF minus the ACF of the synthesized beam} 
\label{fig:fig4}
\end{figure}

For convenience, by replacing the sky position of (x,y) with the pixel position \textbf{x}, noise map of the image in Eq.~(\ref{eq:eq8}) is written as
\begin{equation}
    N(\mathbf{x})=b(\mathbf{x})*\hat{N}(\mathbf{x})=\sum_{i,j}b(\mathbf{x}_{i,j})\hat{N}(\mathbf{x}+\mathbf{x}_{i,j}).
\end{equation}
The autocorrelation of the noise map is\cite{Refregier1998-ui}
\begin{equation}
\begin{split}
        \xi(\mathbf{x}_{i,j})&=\langle N(\mathbf{x}+\mathbf{x}_{i,j})N(\mathbf{x})\rangle \\
        &=\langle \sum_{i',j'}b(\mathbf{x}_{i',j'})\hat{N}(\mathbf{x}+\mathbf{x}_{i,j}+\mathbf{x}_{i',j'})\sum_{i'',j''}b(\mathbf{x}_{i'',j''}\hat{N}(\mathbf{x}+\mathbf{x}_{i'',j''}) \rangle \\
        &=\sum_{i',j'}\sum_{i'',j''}b(\mathbf{x}_{i',j'})b(\mathbf{x}_{i'',j''})\langle\hat{N}(\mathbf{x}+\mathbf{x}_{i,j}+\mathbf{x}_{i',j'})\hat{N}(\mathbf{x}+\mathbf{x}_{i'',j''}) \rangle\\
        &=\sigma_N^2 \alpha(\mathbf{x}),
\end{split}\label{eq:eq10}
\end{equation}
with beam autocorrelation $\alpha(\mathbf{x}_{i,j})$
\begin{equation}
    \alpha(\mathbf{x}_{i,j}) = \sum_{i',j'}b(\mathbf{x}_{i',j'})b(\mathbf{x}_{i,j}+\mathbf{x}_{i',j'}),
\end{equation}
where we used the noise ACF property of uncorrelated noise $\hat{N}$ (Eq.~\ref{eq:eq3}) for the fourth equality. Equation~\ref{eq:eq10} implies that the noise ACF is related to the ACF of the synthesized beam with a constant multiplicative factor, which is the variance of the noise. In Fig.~\ref{fig:fig4} we compare the ACF of noise and that of the synthesized beam, along with the residuals (noise ACF minus beam ACF). Although the noise ACF and beam ACF show a common characteristic pattern, they do not completely coincide, showing an extended weak positive correlation and a relatively large negative around the main beam in the residual. This disagreement is likely due to not only (1) the remaining contamination from the emission from the sources and background undetected sources, but also (2) the process involved in the imaging process. These are discussed in detail in the Appendix~\ref{sec:ap1} comparing Fig.~\ref{fig:fig4} obtained from the actual data with the one (Fig.~\ref{fig:fig12}) obtained from the simulated data with a similar observational setup and realistic noise in visibility but without emissions in the sky.

Our interest is on the statistical property of the noise in the image plane, which we characterize by the noise ACF including effects of contaminations from the source and the imaging process. Note that Davis et al. 2017\cite{Davis2017-lu} used the covariance matrix computed from the synthesized pattern rather than the actual noise in the image to consider the effect of the interpixel correlation of the noise on the fitting. Given that the synthesized beam pattern differs from the actual noise correlation as shown, a noise ACF, which truly characterizes the noise, is recommended to use. 

Due to the limited spatial frequency coverage, the synthesized beam $b(x,y)$ has a complex structure with sidelobes that extend from the center to a large radius. The flux from the source is spread out by the side lobes to the distant pixels in the image. The clean algorithm, which is most commonly used in radio imaging, deconvolves the beam pattern $b(x,y)$ for signals with high S/N ($S(x,y)/\sigma_\mathrm{N})>1.5-3$) and replaces it with a clean beam without side lobes (a Gaussian that approximates the mainlobe of the synthesized beam). The clean algorithm successfully eliminates the influence of the sidelobe and produces a high-fidelity image. However, the spatial correlations existing in stochastic noise $N(x,y)$ cannot be removed by the clean algorithm. Therefore, it is important to evaluate their effects on image analysis and signal detection, which we will describe in the next Section~\ref{sec:result2}. 

\section{EXAMPLE APPLICATION TO SCIENTIC DATA}
\label{sec:result2}
\subsection{Contribution of the correlated noise to the statistical uncertainty in the measured flux}

The most fundamental measurement of astronomy is the total flux spreading over some sky region of interest in the images, which are measured by summing the pixel values over the region of interest (i.e., aperture photometry in optical astronomy, so we call pixel regions to be integrated as an aperture in this paper). In particular, at the submillimeter band of ALMA, the flux of the continuum emission mainly arising from dust, and line emission and absorption by the various atomic and molecular gases are used to estimate the physical properties of the interstellar medium (e.g., dust mass, gas mass, the energy source of the ionization or excitation, etc.). Therefore, it is important to estimate the uncertainty of the measured quantities. As shown in Fig. 2, the noise in interferometric images significantly correlates between pixels, making the estimation of the noise in the integrated flux difficult. The previous literature estimates the statistical uncertainty of the integrated flux by two methods: (1) randomly placing apertures, which have the same shape used for measuring the flux, in the noise region of the image, measuring the sum within the aperture and then adopting the rms as the noise in the integrated flux \cite{Harikane2020-nr}; and (2) assuming that the regions in the image separated with a beam size do not correlate and adopting $\sigma_{\mathrm{N}}\sqrt{N_\mathrm{beam}}$, where $\sqrt{N_\mathrm{beam}}$ is the number of beams (independent regions) in the aperture. $N_\mathrm{beam}$ is estimated as $A_{\mathrm{aperture}}/A_{\mathrm{beam}}$, where $A_{\mathrm{aperture}}$ and $A_{\mathrm{beam}}$ are the aperture area and the clean beam area, respectively \cite{Alatalo2013-ix}. For convenience, we call methods (1) and (2) "random aperture method" and "independent beam method," respectively, in this paper. This section introduces how to derive the statistical uncertainty associated with the spatially integrated flux directly from the computed noise ACF.

\begin{figure*}[h]
\begin{minipage}[t]{0.55\linewidth}
    \includegraphics[width=\linewidth]{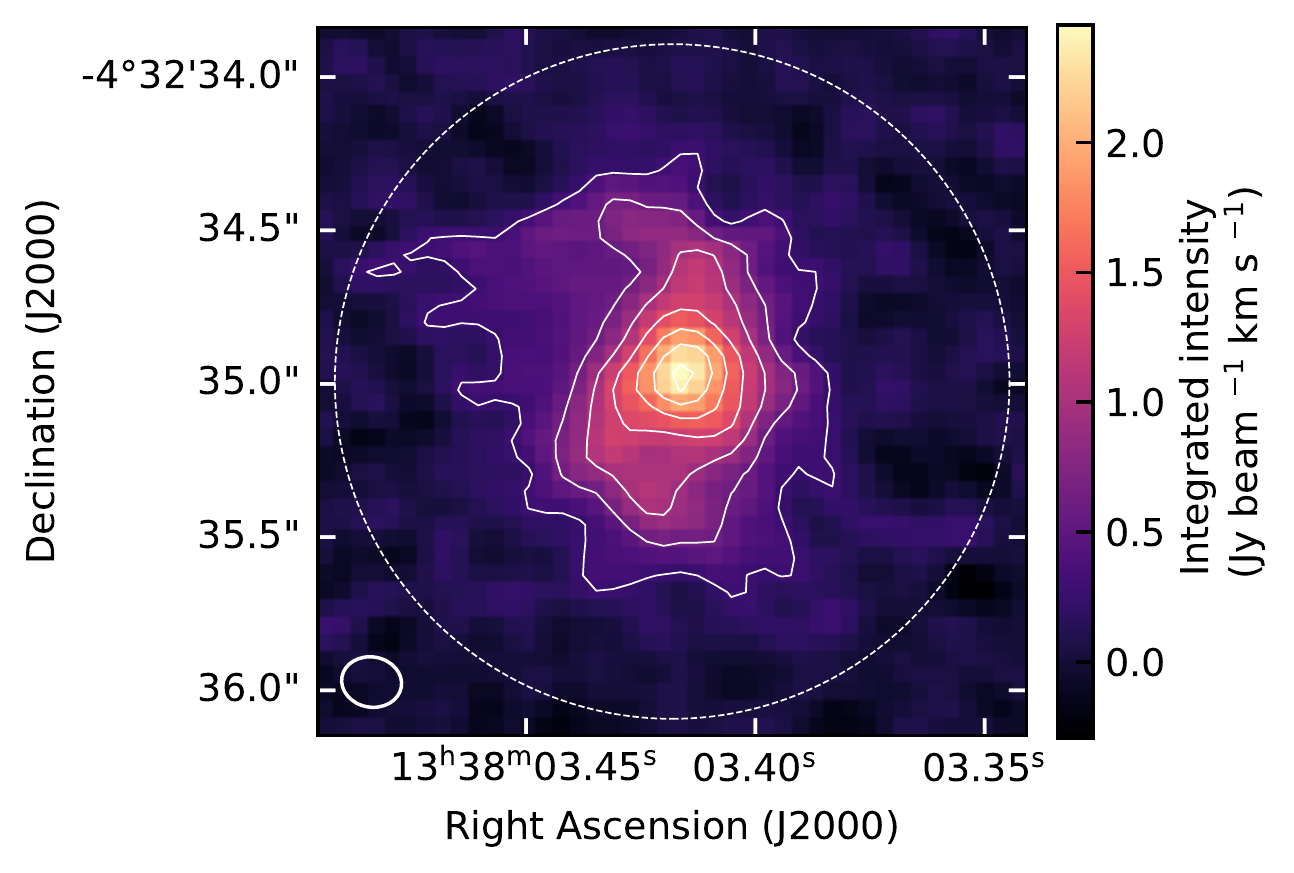}
    \caption{[C\textsc{ii}] velocity integrated intensity map of BRI 1335-0417\cite{Tsukui2021-sg} The white contour is shown every 4$\sigma$ from 3$\sigma$ to 27$\sigma$. The white elipse shown in bottom-left corner indicates the FWHM of the main lobe of the synthesized beam. The dotted line circle shows the aperture corresponding to the point with the largest number of pixels shown in Fig \ref{fig:fig6}. The [CII] spectrum within the aperture is also shown in Fig \ref{fig:fig9}. \\\label{fig:fig5}}
\end{minipage}
\hspace{0.01\textwidth}
\begin{minipage}[t]{0.45\linewidth}
    \includegraphics[width=\linewidth]{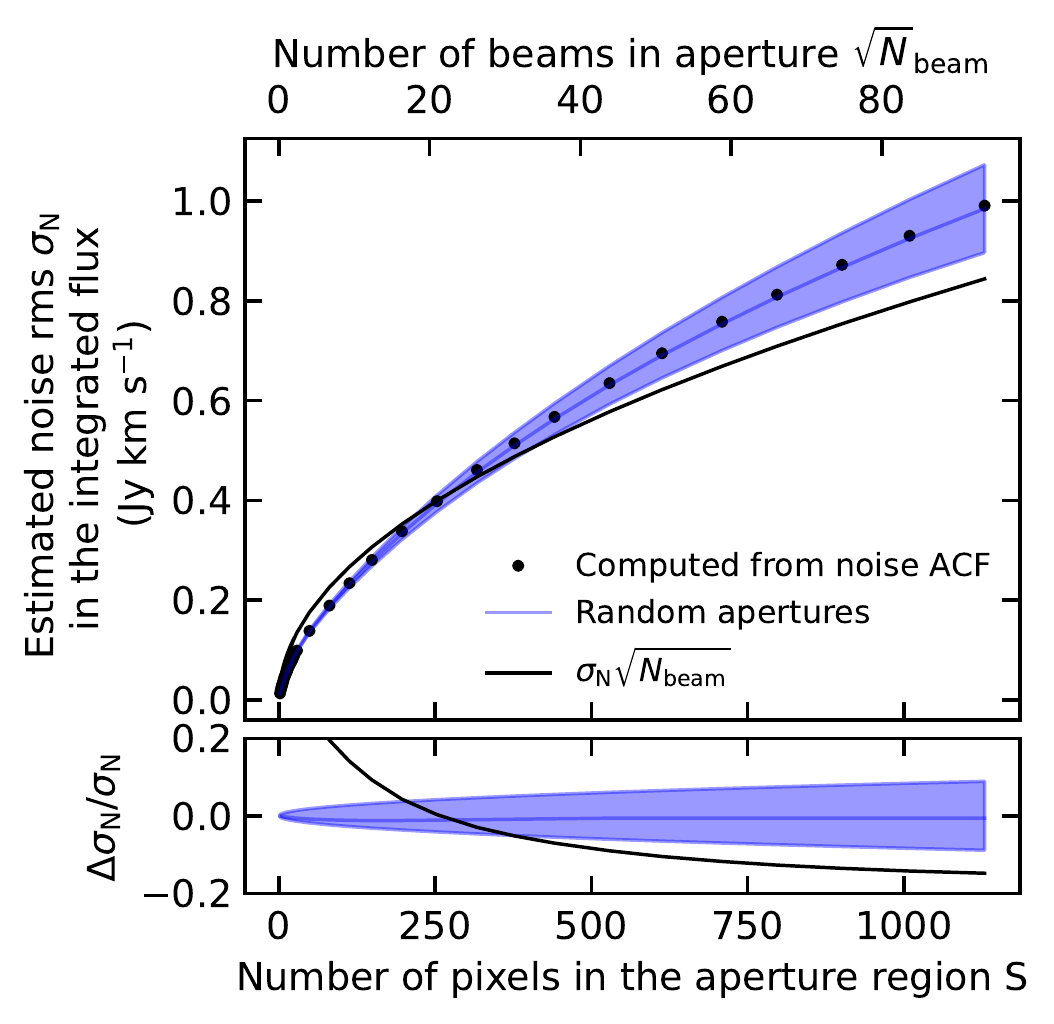}
    \caption{Top: The noise rms in the integrated flux for apertures with different sizes estimated by various methods: the one computed from the noise ACF (black points), the one estimated from the random aperture methods (blue line) with the standard error (blue shade), and the one estimated from the independent beam method ($\sigma_{\mathrm{N}}\sqrt{N_\mathrm{beam}}$). Bottom: the fractional difference between the one computed from noise ACF and the ones from the random aperture method and independent beam method. \\\label{fig:fig6}}
\end{minipage}
\end{figure*}

We consider adding all the pixel values at pixel positions \{$\mathbf{x}$\} within the sky region of interest S. The random noise $N(\mathbf{x})$ in the map is characterized by the noise ACF, $\xi(\mathbf{x}_{i,j})$. The 1$\sigma$ statistical uncertainty associated with the summed value within the pixel reagion S, $\sigma_{\mathrm{int}}$, can be estimated as
\begin{equation}
\begin{split}
    \sigma_{\mathrm{int}}^2 & = \mathrm{Var}(\sum_{\mathbf{x}<S}N(\mathbf{x}))\\
     & = \sum_{\mathbf{x}<S}\mathrm{Var}(N(\mathbf{x}))+\sum_{\mathbf{x}<S} \sum_{\substack{ \mathbf{x}'\neq\mathbf{x}\\\mathbf{x}'<S} }\mathrm{Cov}(N(\mathbf{x}),N(\mathbf{x}'))\\
     & = N_{\mathrm{pix}}\sigma_\mathrm{N}^2+\sum_{\substack{\mathbf{x}_{i,j}=\mathbf{x}-\mathbf{x}'\\ \mathbf{x}'\neq\mathbf{x}\\\mathbf{x}, \mathbf{x}'<S} }\xi(\mathbf{x}_{i,j}),
\end{split}\label{eq:14}
\end{equation}
where $N_{\mathrm{pix}}$ is the number of pixels in the region S. Var and Cov indicate the variance and covariance, respectively. The second term of the last line is the sum of the noise ACF for all possible pixel separation vectors $\mathbf{x}_{i,j}$ between two pixels within the reagion S. If the noise does not have inter-pixel correlation, the second term becomes zero, resulting in $\sigma_{\mathrm{int}}=\sigma_{\mathrm{N}}\sqrt{N_{\mathrm{pix}}}$. As an illustration, we used the spatially resolved [C\textsc{ii}] moment 0 map of BRI1335-0417 taken by ALMA, shown in Fig.~\ref{fig:fig5}, where the emission spreads over multiple pixels. 
We calculated the noise in the integrated flux measured using a variety of apertures S with different sizes. The largest aperture is a dotted line shown in Fig.~\ref{fig:fig5}.
Figure~\ref{fig:fig6} shows the computed noise from the measured noise ACF compared to the previously used "random aperture" and "independent beam" ($\sigma_{\mathrm{N}}\sqrt{N_\mathrm{beam}}$) methods. The noise calculated from the ACF is in excellent agreement with the random aperture method, while the independent beam method tends to overestimate at smaller apertures and underestimate at large apertures, showing that the assumption of "independent beam" is oversimplified. 
When the field of view is small (the field of view becomes smaller at higher frequency bands in ALMA) or the aperture area becomes larger, the random aperture method cannot place apertures randomly in the limited area of the emission free region, and thus the standard error of the estimate increases, as shown in the blue shade in Fig.~\ref{fig:fig6}. However, the proposed method can provide the best estimate by exploiting all available data to estimate noise ACF. 

\begin{figure*}[h]
\begin{minipage}[t]{0.49\linewidth}
    \includegraphics[width=\linewidth]{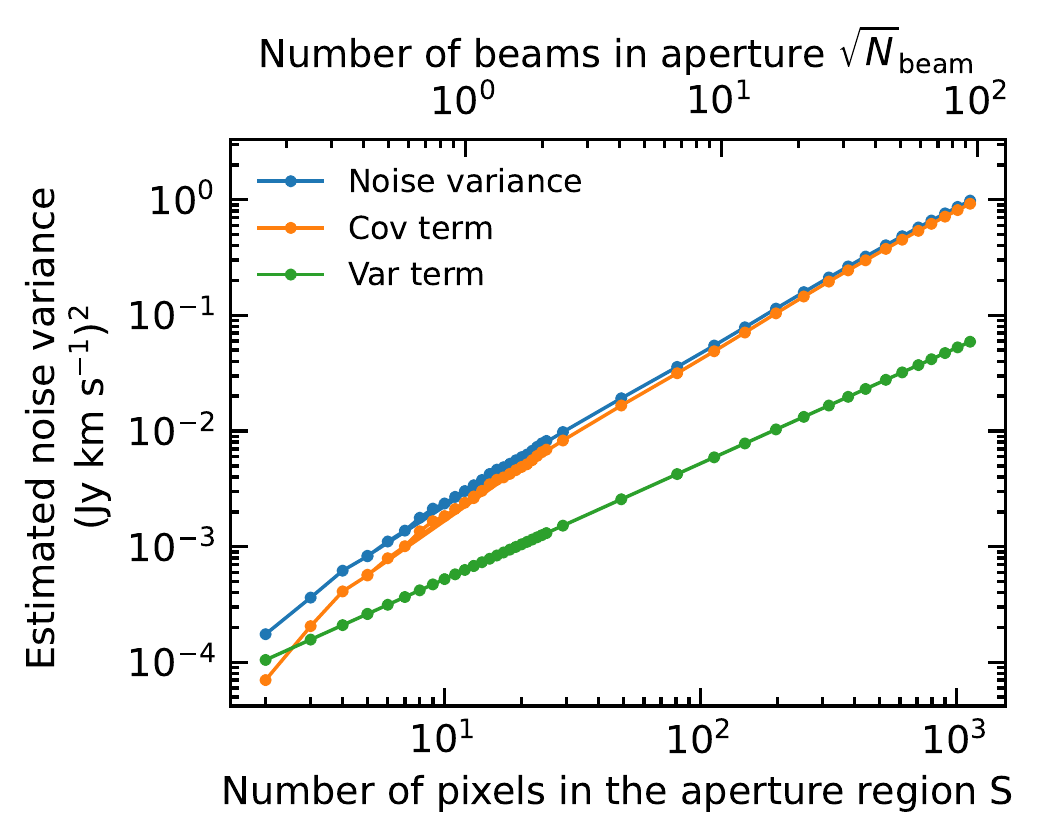}
    \caption{The measured noise variance estimated from the noise ACF for different appertures S (blue dot and line), with the contribution from the noise variance of the individual pixels in the apperture (the first term of the last line in Eq.~\ref{eq:14}, green dot and line) and the contribution from the covariance due to the interpixel correlation (the second term of the last line in Eq.~\ref{eq:14}, orange dot and line) \label{fig:fig7}}
\end{minipage}
\hspace{0.01\textwidth}
\begin{minipage}[t]{0.49\linewidth}
    \includegraphics[width=\linewidth]{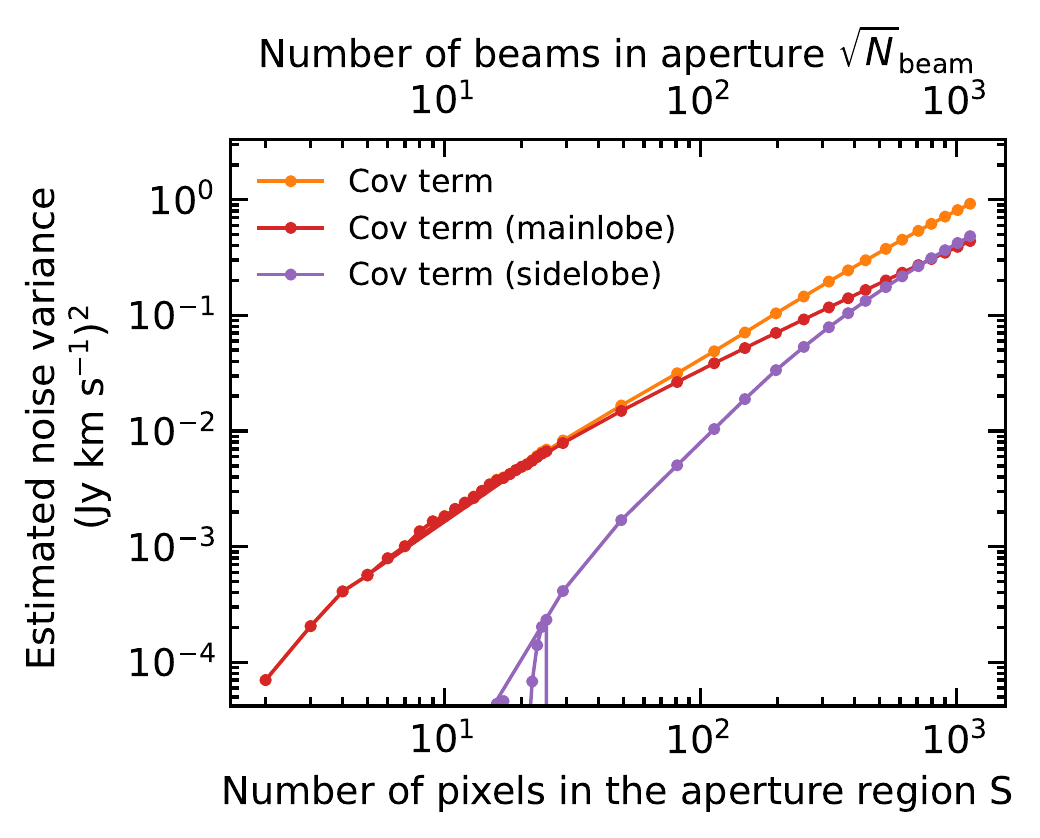}
    \caption{The covariance term due to the interpixel correlation shown in Fig.~\ref{fig:fig7}. (the second term of the last line in Eq.~\ref{eq:14}, orange dot and line), which are further decomposed into the component of the long range interpixel correlation due to the sidelobe of the synthesized beam and short range interpixel correlation due to the mainlobe of the beam. \label{fig:fig8}}
\end{minipage}
\end{figure*}

To demonstrate the significant effect of spatially correlated noise, Figure~\ref{fig:fig7} shows the noise variance $\sigma_\mathrm{N}^2$ in the integrated flux calculated from the ACF along with the contributions of the noise variance of individual pixels in the aperture (the first term in Eq.~\ref{eq:14}, this is the value we obtain if we are unaware of the noise correlation) and the interpixel correlation in the aperture (the second term in \ref{eq:14}). In the case of an aperture with two pixels, it is not mathematically allowed for the first term to exceed the second term. However, as the number of pixels in the aperture, $N_{\mathrm{pix}}$, increases, the contribution of the second term becomes dominant because the number of first terms is $N_{\mathrm{pix}}$ while the number of second terms is $N_{\mathrm{pix}}(N_{\mathrm{pix}}-1)$. The ignorance of the noise correlation will lead to a significant underestimation of the integrated flux uncertainty. Figure~\ref{fig:fig8} further divides the variance due to noise correlation into the effects of the mainlobe (correlation due to the mainlobe of the synthesized beam) and the sidelobe (long-range pixel correlation due to the sidelobe of the synthesized beam). To compute these components separately, we define the noise ACF inside/outside of the beam FWHM as short-range/long-range correlation components, respectively. After $N_{\mathrm{pix}}$ exceeds 200, the effect of the sidelobe becomes significant. This explains the deviation of the estimate by $\sigma_{\mathrm{N}}\sqrt{N_\mathrm{beam}}$ from the value derived by other methods shown in Fig.~\ref{fig:fig6}; the number of independent beams $N_\mathrm{beam}$ is estimated by the area of the clean beam, and the long-range correlation due to the sidelobe is not properly taken into account. The total [C\textsc{ii}] flux measured with the aperture shown in Fig.~\ref{fig:fig5} is 29.51 $\pm$ 1.05 Jy km s$^{-1}$ (1$\sigma$ statistical uncertainty calculated from the noise ACF).

\begin{figure}[h]
\centering
\includegraphics[width=0.5\textwidth]{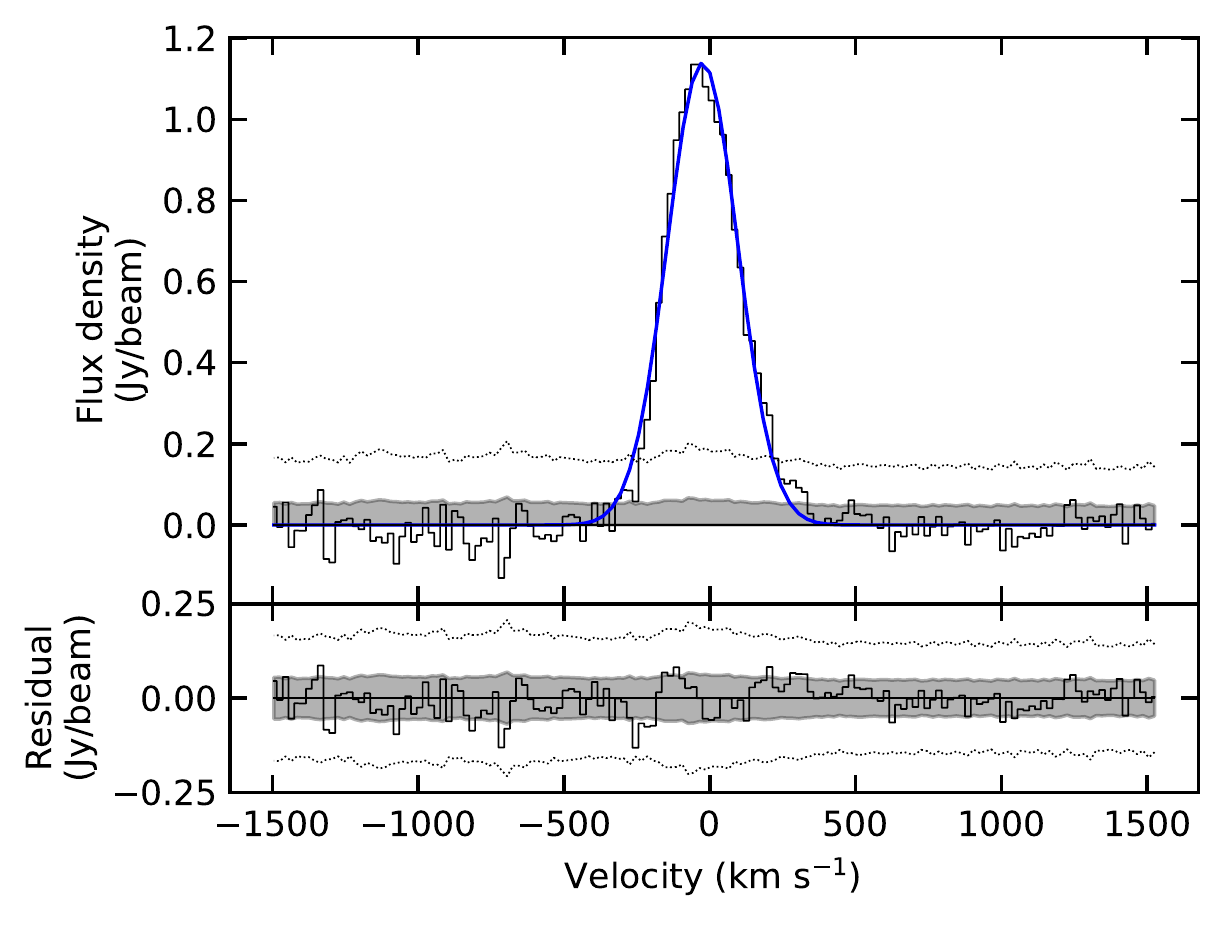}
\caption{The top panel shows the spatially integrated flux within the aperture shown in Fig.~\ref{fig:fig5} (solid black line) and the best-fit Gaussian (solid blue line). The bottom panel shows the residual (measured spectrum minus the best-fit Gaussian). The gray shade and dashed line in both panels show the statistical uncertainty of 1$\sigma$ and 3$\sigma$, respectively, computed from the noise ACF. The figure is reproduced from Tsukui and Iguchi \cite{Tsukui2021-sg} where the underlying noise is recalculated by the noise ACF of each velocity channel. \label{fig:fig9}} 
\end{figure}

Another important measurement in astronomy is the shape of the spectrum integrated over a certain region of interest. Similarly to deriving the noise in the integrated flux, we can derive the underlying noise in the spatially integrated spectrum using the noise ACFs computed for every velocity channel of the data cube. Fig.~\ref{fig:fig9} shows the spatially integrated [CII] spectrum enclosed by the aperture shown in Fig.~\ref{fig:fig5} with 1$\sigma$ and 3$\sigma$ noise levels. The [CII] spectrum of BRI1335-0417 is well described by a single Gaussian without deviations from the Gaussian above 3$\sigma$. Because emission lines from astronomical interesting phenomena such as outflows, tidal tails etc. are faint, it is important to accurately estimate the noise, otherwise it will lead to false detections.

\subsection{Simulating the noise maps}
In image analysis, generating random noise based on the statistical properties of the noise is useful for evaluating the significance of the results. In this section, we describe how to generate random noise based on the noise ACF, which fully characterizes the correlated Gaussian noises, and demonstrate the ingnorance of the noise correlation leading to the misinterpretation of the result with the example analysis. 

Once the noise ACF is measured, the noise at the positions of the $i,j$ pixels $\mathbf{x}_{i,j}$ in the images with size of $M\times M$ pixels can be generated randomly by the joint probability distribution, the probability that $N(\mathbf{x}_{i,j})$ takes the value in the small intervals ($N_{i,j}+\mathrm{d}N_{i,j}$) given by\cite{Binney2008-fu}
\begin{equation}\label{eq:jointprob}
    \mathrm{d}p=\frac{\mathrm{d}N_{1,1}\cdot\cdot\cdot\mathrm{d}N_{i,j}\cdot\cdot\cdot\mathrm{d}N_{M,M}}{(2\pi)^{M^2/2}|\mathbf{C}|^{1/2}}\exp{\left(-\frac{1}{2}\Sigma^M_{a,b,c,d=1}N_{a,b}\mathbf{C}^{-1}_{a-c,b-d}N_{c,d}\right)},
\end{equation}
where $\mathbf{C}^{-1}$ is the inverse of the matrix $\mathbf{C}$ defined by
\begin{equation}
    C_{i,j}=\xi(\mathbf{x}_{i,j})
\end{equation}

\begin{figure}[h]
\centering
\includegraphics[width=0.43\textwidth]{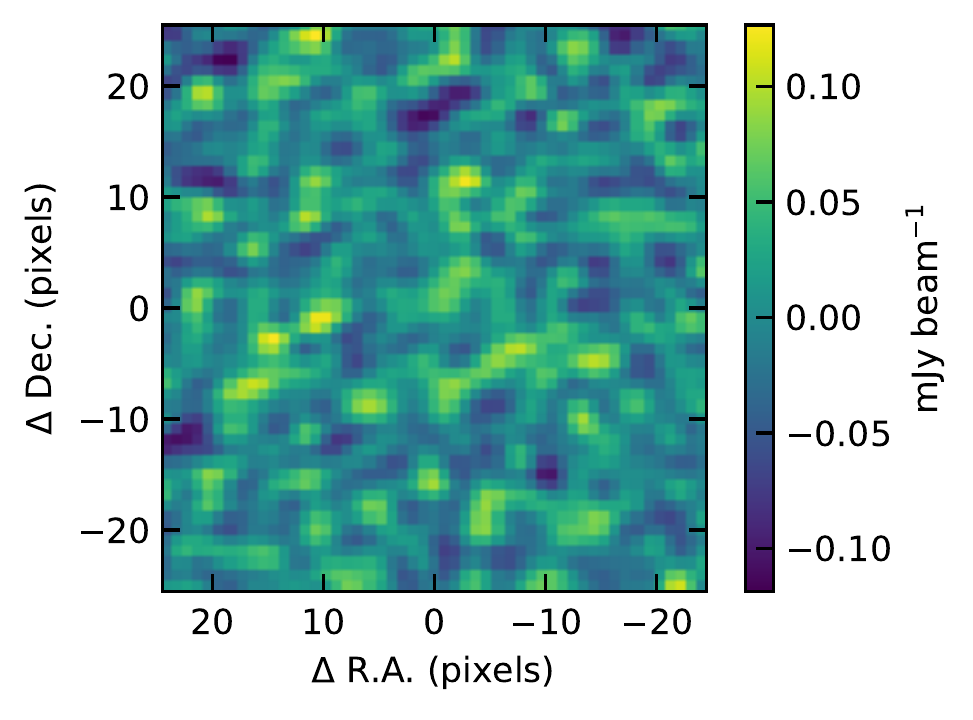}
\includegraphics[width=0.43\textwidth]{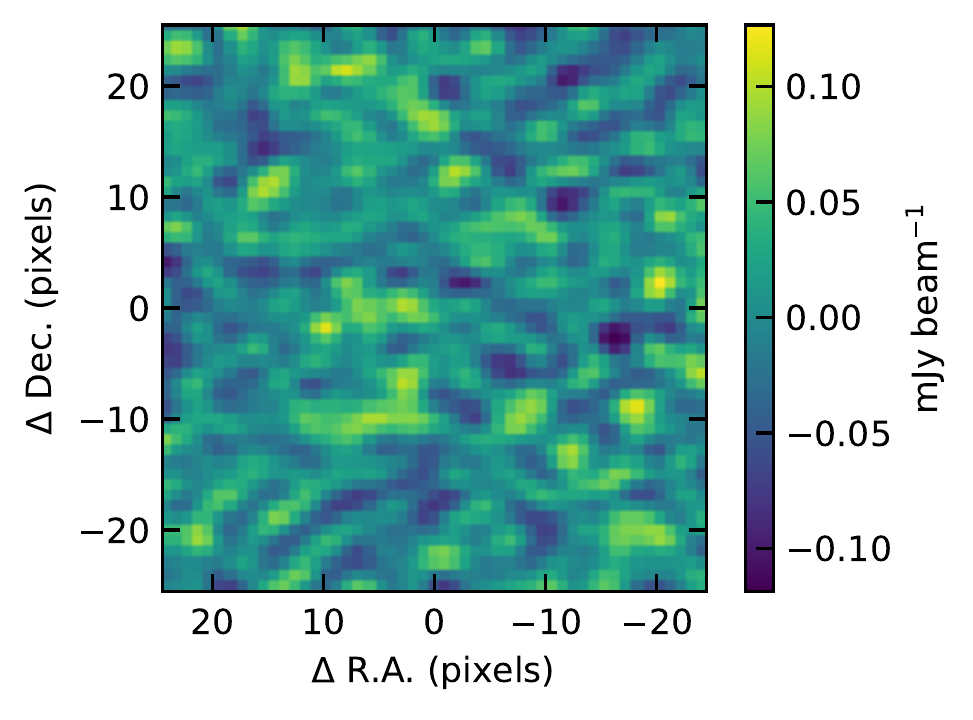}
\includegraphics[width=0.43\textwidth]{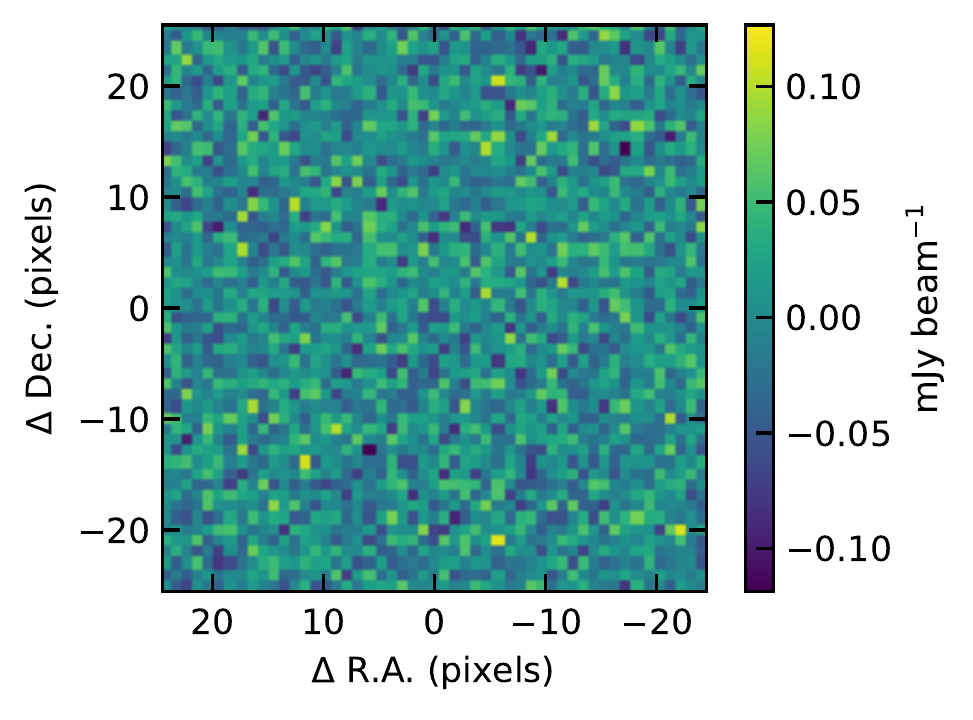}

\caption{Noise map in the observed data (top left), noise map generated from the measured noise ACF (top right), and uncorrelated noise map (bottom). All of which have the same standard deviation $\sigma_{N}$. \label{fig:fig10}} 
\end{figure}

Figure~\ref{fig:fig10} shows the comparison of the noise of the observed data, the noise randomly generated from the measured noise ACF using the joint Gaussian probability distribution (Eq.~\ref{eq:jointprob}; we used the \textsc{multivariate}\_\textsc{normal} function from the scipy package\cite{2020SciPy-NMeth}) and the spatially uncorrelated Gaussian noise, all of which have the same standard deviation. The observed noise and the randomly generated noise using noise ACF are qualitatively
similar, while the spatially correlated noise and the spatially uncorrelated noise look completely different, illustrating how dangerous it is to assume naively uncorrelated noise in the image analysis.

\begin{figure}[h]
\centering
\includegraphics[width=0.43\textwidth]{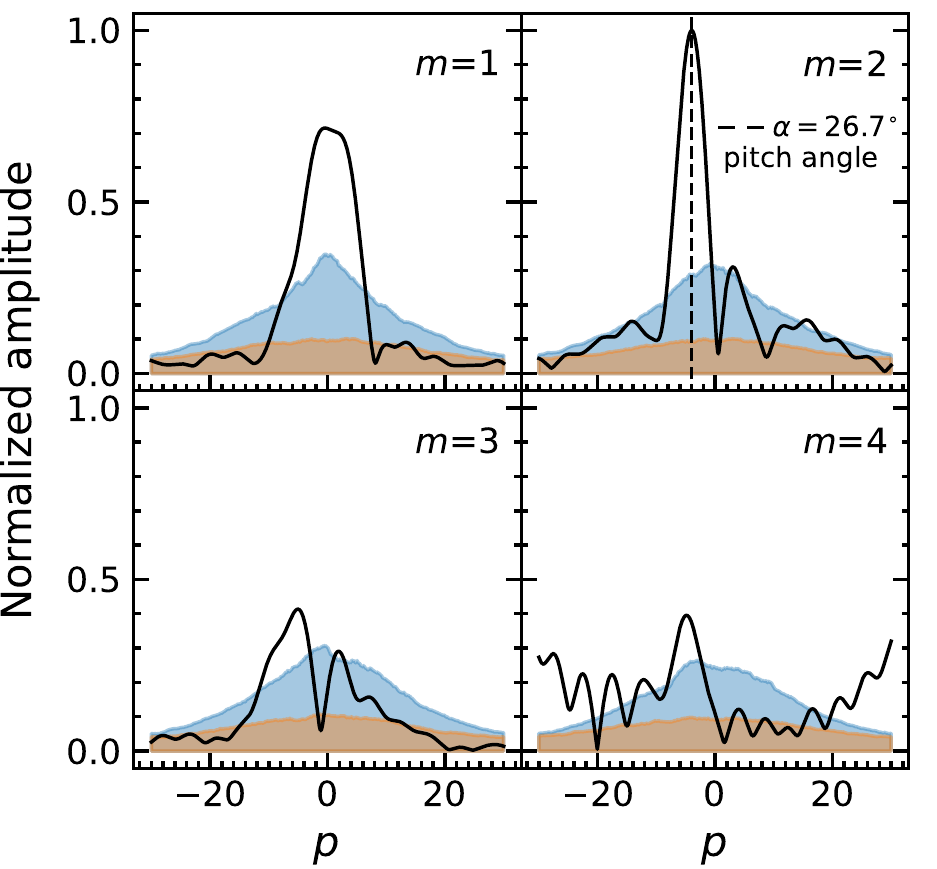}
\caption{Fourier spectra of logarithimic spiral models (solid black lines) and the underlying 1 $\sigma$ noise due to the noise in the images computed from the simulated noise maps by the noise ACF (blue shaded region) and the white Gaussian noise maps with the same standard deviation (orange shaded region). The peak in $m=2$ indicates that the dominant component of the image is a 2 armed spiral structure with a pitch angle of $\alpha=26.7$. The figure is adopted from Tsukui and Iguchi \cite{Tsukui2021-sg} and the underlying noise spectrum is recalculated using the noise maps simulated by the noise ACFs. \label{fig:fig11}} 
\end{figure}

In the literature, emission-free regions have been extracted and used as realistic noise maps to estimate the statistical uncertainty associated with the parameters derived by the model fitting \cite{Boizelle2019-wg}. The field of view of the interferometric observation is generally small, preventing us from obtaining a sufficient number of independent noise maps to conduct Monte Carlo experiments. The proposed method can generate random noise repeatedly from the measured statistical properties of the noise (the noise ACF) in the interferometric image with the best precision limited by the available area of the emission-free region. Although there are many potential applications of generating noise, we illustrate the significance of the correlated noise using only one example: the image analysis done in Tsukui and Iguchi 2021, which expand the image with series of logarithmic spirals and identify the dominant spiral structure in the image. In Fig.~\ref{fig:fig11}, we showed the Fourier spectrum of the logarithmic spirals in the [CII] intensity map of the BRI1335-0417 shown in Fig.~\ref{fig:fig5} (black solid line) and the underlying noise spectrums estimated from the simulated noise maps from noise ACF (blue shaded region) and the simulated noise maps with no spatial correlation but the same standerd deviation $\sigma_{\mathrm{N}}$ (orange shaded region). The spectrum shows the amplitude of the logarithmic spiral with $m$ arms ($m$-fold symmetry) with the pitch angle of $\alpha=\arctan{(-m/p)}$, which was calculated as follows. The image (Fig.~\ref{fig:fig5}) was first deprojected to be viewed face-on with an inclination angle of 37.8$^{\circ}$ and a position angle of 4.5$^{\circ}$, where the package \textsc{scikit-image} performed rotation and stretching for the deprojection (which induces the additional noise correlation in the image). Then the amplitude of each Fourier component of the logarithmic spiral $m$ and $\alpha$ is calculated (see equation S2 in Tsukui and Iguchi 2021 \cite{Tsukui2021-sg}). The noise spectrum is measured by repeatedly generating 300 noise maps, calculating the amplitude of each Fourier component in the same way as the image, and taking the 84th percentiles. The noise spectrum computed by assuming the white Gaussian noise map significantly underestimates the true noise spectrum, which could lead to the false detection of statistically insignificant structures such as the second peak in $m=2$ and multiple peaks in $m=3, 4$. The estimated noise spectra are higher than those reported in the original paper because the image stretching and rotation were not applied to the noise maps in the original paper, resulting in an underestimation of the noise level. However, the result of the original article does not change except for the updated statistical significance of each peak in $m=1,2,3,4$ being 2.1$\sigma$, 3.5$\sigma$, 1.7$\sigma$, and 1.6$\sigma$, respectively. $m=1$ peak with 2.1$\sigma$ corresponds to the fact that the northern arm is longer than the southern arm.

\section{CONCLUSIONS}
Understanding interpixel correlation of noise in interferometric images is important to correctly estimate the uncertainty of the data analysis. We show that the noise ACF of ALMA noise image has a pattern similar to that of the synthesized beam (dirty beam) and that the spatial correlation of the noise originates from the limited uv coverage or synthesized beam pattern. We present a method for calculating the statistical uncertainty in spatially integrated quantities such as flux and spectrum directly from the noise ACF. The contribution of the spatial correlation of the noise dominates in the estimated statistical uncertainties as the aperture size increases. Also, we present a method for simulating the noise from the noise ACF for Monte Carlo experiments. We demonstrate the example applications to the scientific data showing that the ignorance of noise correlation lead to significant underestimation of statistical uncertainty of the results. We made a Python code publicly available \url{https://github.com/takafumi291/ESSENCE}. 

\appendix 
\section{The difference between\\ the noise ACF and the synthesized beam ACF}
\label{sec:ap1}
\begin{figure}[h]
\centering
\includegraphics[width=0.43\textwidth]{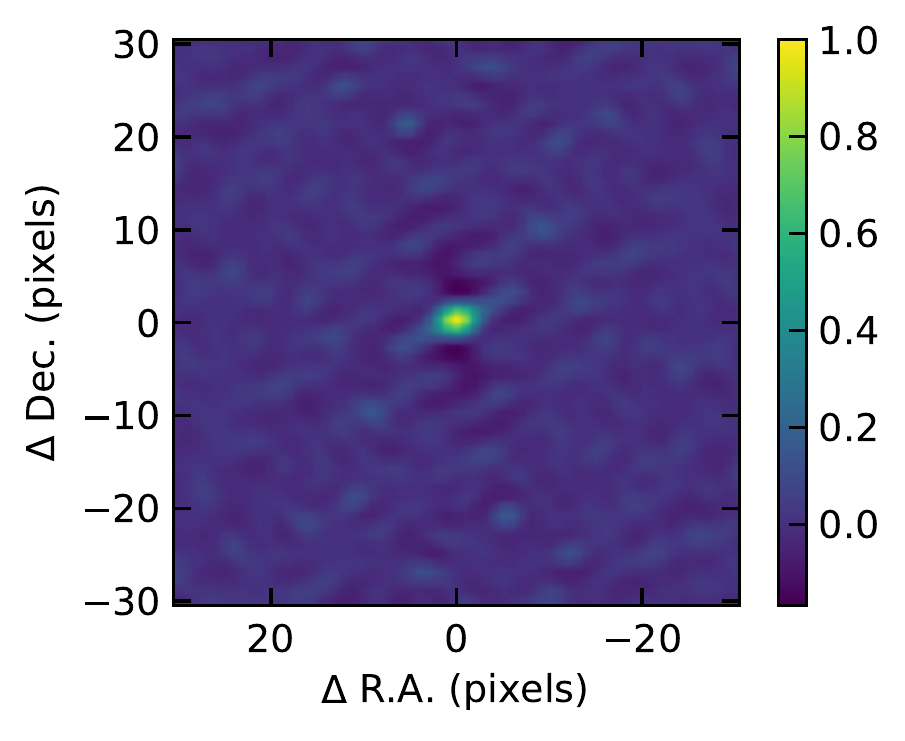}
\includegraphics[width=0.43\textwidth]{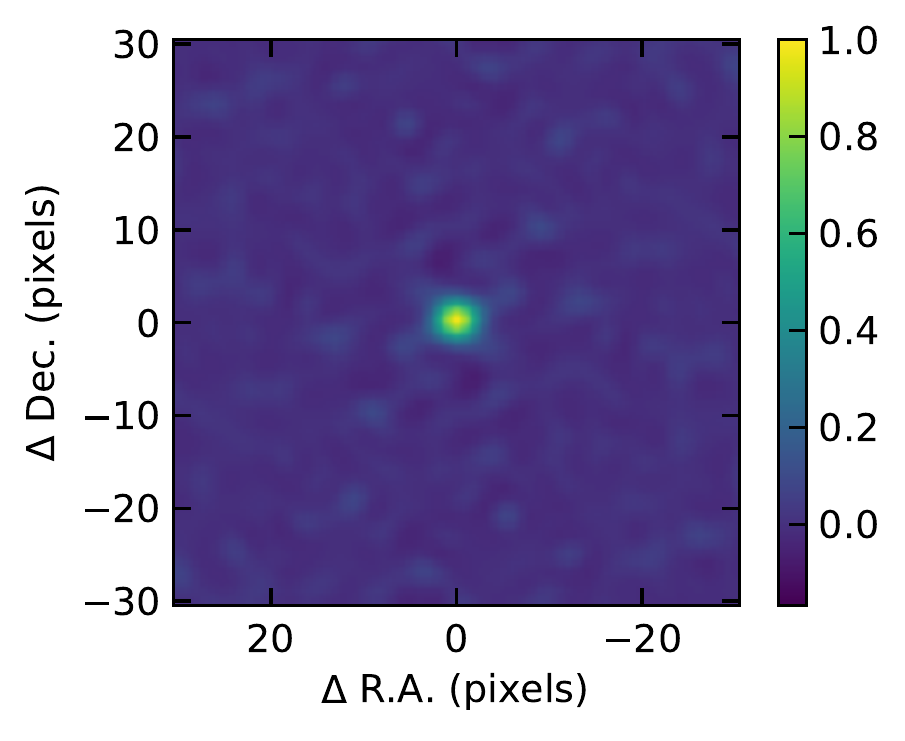}
\includegraphics[width=0.43\textwidth]{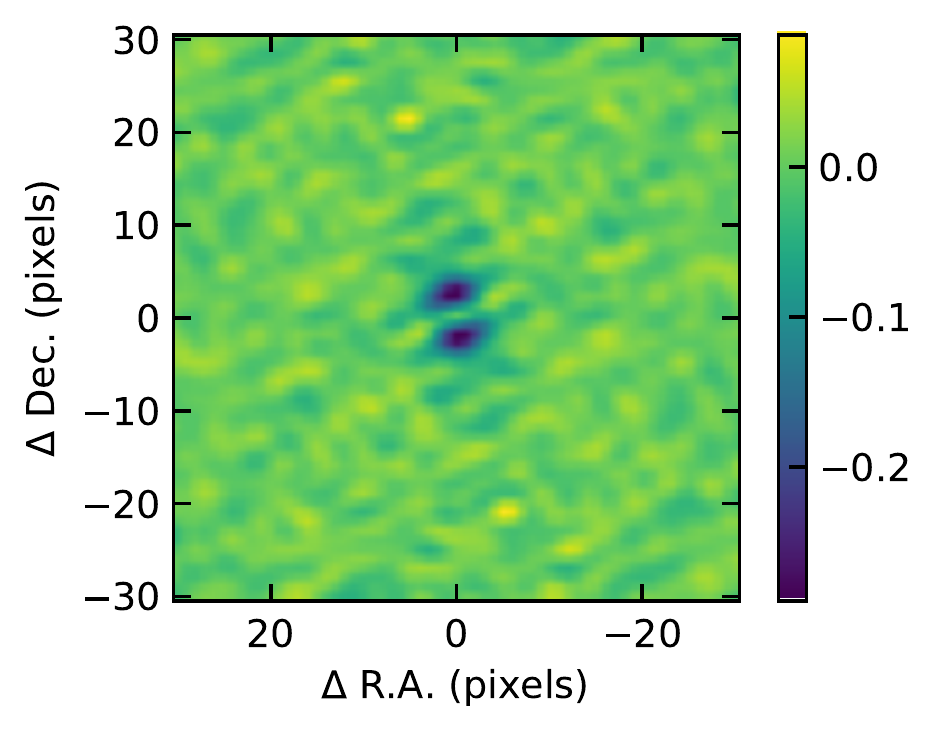}
\caption{Same as Fig.~\ref{fig:fig4} but for the noise image produced from the simulated data with a similar observational setup and without sky emission.} 
\label{fig:fig12}
\end{figure}

\begin{figure}[h]
\centering
\includegraphics[width=0.43\textwidth]{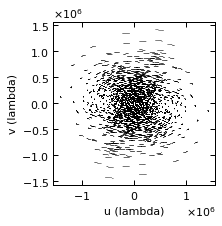}
\caption{The spatial transfer function of the simulated observation $W(u, v)$} 
\label{fig:fig13}
\end{figure}

Equation \ref{eq:eq10} implies that the noise ACF is identical to the ACF of the synthesized beam with a constant multiplicative factor. However, they do not completely coincide, showing an extended weak positive correlation and a relatively large negative around the main beam in the residual (noise ACF - ACF of the synthesized beam, Fig.~\ref{fig:fig4}). To investigate the origin of the feature, we simulate the observation with a similar setup: hour angle, sky position, antenna configuration, and realistic atomospheric noise, but without emissions in the sky, using \textsc{simalma} in \textsc{casa}\cite{McMullin2007-bg}\footnote{The actual data is taken with $\sim$2 hour observation, with 1 hour of integration on a source intermittently separated by calibrator observations and other overheads. In comparison, the simulated observation is a continuous 1 hour of integration on the source.}. The visibility obtained is imaged with the same imaging parameters as those used for the actual data. Fig.~\ref{fig:fig12} shows the noise ACF, the ACF of the synthesized beam, and their residual (noise ACF - ACF of the synthesized beam) for the simulated observation without sky emisssion. The residual does not show the extended positive correlation pattern seen in Fig.~\ref{fig:fig4}, suggesting that the positive correlation pattern may be due to the sky emission of the sources and undetected background sources. The emissions from the astronomical sources are spread in the noise region by long-range sidelobes of the synthesized beam, which are generally removed for bright emissions down to 1.5 to 3 sigma by \textsc{clean}. Weaker emissions and its leakage into the surrounding pixels remain in the noise region. These un\textsc{clean}ed components and weak background sources hidden in the noise map may produce the faint positive correlation pattern in the residual Fig.~\ref{fig:fig4}.
However, the residual obtained for the simulated observation without sky emission (\ref{fig:fig12}) still shows a similar negative around the main beam as seen in Fig.~\ref{fig:fig4}, suggesting that the large negative cannot be attributed to emissions from the sky and may be due to the process involved in the imaging. In practice, the Fourier transform in the imaging is performed by the discrete Fourier transform (FFT), in which the visibility data is evaluated discretely as the rectangular grid. The visibility data represented as ($V(u,v)+\hat{N}_{\mathrm{vis}}(u,v))W(u,v)$ in Eq.~\ref{eq:eq8} are interpolated\footnote{more precisely, the visibility data are convolved by a function chosen for minimizing the image ailiasing to produce continuous distribution \cite{Thompson2017-fp}} to estimate the visibility value in the center of the grid. Each grid is further weighted depending on the number of data points in the grid cell or the rms error of the data in the cell\cite{Briggs1995-yt}. The negative feature in the north-south direction around the main beam may correspond to the fact that the sampling of $u,v$ is relatively denser in the north-south direction (see Fig.~\ref{fig:fig13}), which is more affected by the averaging process, than in the east-west direction. 

\section*{ACKNOWLEDGEMENTS}
TT deeply thanks Ma, Yik Ki (Jackie) at Australian National University for providing constructive and insightful feedback on the manuscript. TT also thanks Emily Wisnioski, Daisuke Iono, Jorge A. Zavala, and Izumi Takuma for encouraging comments and helpful discussions. TT was supported by the ALMA Japan Research Grant of NAOJ ALMA Project, NAOJ-ALMA-264. This research has used the following data: ADS/JAO.ALMA \#2017.1.00394.S. ALMA is a partnership of ESO (representing its member states), NSF (USA) and NINS (Japan), together with NRC (Canada), MOST and ASIAA (Taiwan), and KASI (South Korea), in cooperation with the Republic of Chile. The Joint ALMA Observatory is operated by ESO, AUI/NRAO, and NAOJ. Data analysis was carried out in part
on the common-use data analysis computer system at the East Asian ALMA Regional Center (EA ARC) and the Astronomy Data Center (ADC) of the National Astronomical Observatory of Japan (NAOJ). This research made use of \textsc{numpy}\cite{harris2020array}, \textsc{scipy}\cite{2020SciPy-NMeth}, \textsc{matplotlib}\cite{Hunter:2007} , Astropy,\footnote{http://www.astropy.org} a community-developed core Python package for Astronomy \cite{astropy:2013, astropy:2018}, and \textsc{spectral-cube}\cite{2019zndo...2573901G}.

\bibliography{spie.bib} 
\bibliographystyle{spiebib} 

\end{document}